\documentclass[11pt,a4paper]{article}

\usepackage{amsmath,amssymb,amsfonts}
\usepackage{graphicx}
\usepackage{tikz}
\usepackage{cite}

\usepackage[
colorlinks=true,
linkcolor=blue,
citecolor=blue
]{hyperref}

\DeclareMathOperator{\Tr}{Tr}

\title{Radial Convex Geometry of Quantum States and Its Relation to Best Separable Approximation}

\author{
Haonan Qiang\\
\texttt{2120220025@mail.nankai.edu.cn}
}

\date{\today}

\begin{document}

\maketitle

\begin{abstract}

We study bipartite entanglement from the convex geometry of quantum
states. Taking the maximally mixed state as a reference point, we
define a geometric entangled-space quantity
$G(\rho)=[1-L(\rho)]/L(\rho)$ from the relative positions of the
separable and quantum-state boundaries, and introduce a relative
entanglement degree
$Q(\rho)=[(1-p)/p]/G(\rho)$ by comparing it with the robustness
relative to the maximally mixed state.

We apply this geometric construction to the Best Separable
Approximation (BSA). For the optimal decomposition, we derive the
general bound
$(1-p)L_B/[p(1-L_B)]\leq p_0\leq Q(\rho)$ and show that the entangled
component of the BSA has an entangled-space size no smaller than that
of the original mixed state, namely
$[1-L_R]/L_R\leq[1-L_B]/L_B$.

For two-qubit states, the BSA entangled component is a pure
entangled state. Using the PPT criterion, its geometric parameter can
be evaluated explicitly, giving the bound
$(1-p)/(2p)\leq p_0\leq Q(\rho)$. These results provide a simple
geometric description of the relation between BSA, robustness, and the
entangled region of the quantum-state space.

\end{abstract}

%========================================================
%========================================================
\section{Introduction}

Quantum entanglement is an important resource in quantum information
theory. For a bipartite quantum system, the set of separable states
provides a natural reference for describing entanglement\cite{Horodecki2009}. Since the
set of quantum states and the set of separable states both have a
convex structure, their relative geometry provides another way to
describe the position of an entangled state\cite{Zyczkowski1998}.

In this work, we study this geometry by taking the maximally mixed
state as a reference point. An entangled state is connected to the
maximally mixed state by a radial line, and the intersections of this
line with the separable-state boundary and the boundary of the
quantum-state set determine the available entangled region along that
direction. Based on this construction, we introduce a geometric
quantity describing the size of this region and a relative
entanglement degree by comparing it with the robustness with respect
to the maximally mixed state.

We then study the Best Separable Approximation (BSA) of an entangled
mixed state \cite{Lewenstein1998,KarnasLewenstein2001}. For the optimal BSA decomposition, we show that the
separable and entangled components lie on the boundaries of the
separable-state set and the quantum-state set, respectively. More
importantly, the geometric construction gives the relation
\begin{equation}
\boxed{
\frac{1-L_R}{L_R}
\leq
\frac{1-L_B}{L_B},
}
\end{equation}
where $R$ denotes the target state and $B$ denotes the entangled
component in its optimal BSA decomposition. Thus, the entangled
component obtained from the optimal BSA has an entangled-space size no
smaller than that of the original mixed state.

The geometric derivation of this relation does not rely on the PPT
criterion or on the special structure of two-qubit systems. We then
consider the two-qubit case separately, where the remaining BSA
entangled component is a pure state and, for two-qubit systems, the PPT
criterion is necessary and sufficient for separability
\cite{Peres1996,Horodecki1996PPT}. This allows the corresponding
geometric quantities to be evaluated explicitly and gives explicit
bounds relating the BSA weight, the robustness, and the relative
entanglement degree.

The remainder of this paper is organized as follows. In Sec.~II, we
introduce the convex-geometric description of bipartite quantum states
and the associated separable-state set. In Sec.~III, we define the
geometric quantities $G(\rho)$ and $Q(\rho)$ and discuss their
relation to the robustness with respect to the maximally mixed state.
In Sec.~IV, we introduce the BSA decomposition and show that its
separable and entangled components lie on the boundaries of the
corresponding convex sets. In Sec.~V, we construct the geometric
relations for the BSA and derive the corresponding bounds and the
ordering relation between the entangled-space sizes. In Sec.~VI, we
consider two-qubit states, where the PPT criterion allows the
geometric quantities to be evaluated explicitly and gives the resulting
BSA bounds. Finally, Sec.~VII summarizes the main results and discusses
possible extensions.

%========================================================
\section{Convex Geometry of Bipartite Quantum State Space}
%========================================================

Consider a bipartite quantum system
$\mathcal{H}_A\otimes\mathcal{H}_B$ with
$\dim\mathcal{H}_A=d_A$ and $\dim\mathcal{H}_B=d_B$.
Let
\[
d=d_A d_B .
\]
A quantum state is described by a density operator $\rho$ satisfying

\begin{equation}
\rho=\rho^\dagger,\qquad
\rho\geq0,\qquad
\Tr(\rho)=1 .
\end{equation}

A $d\times d$ Hermitian matrix contains $d^2$ independent real
parameters. The trace condition removes one degree of freedom.
Therefore, the space of normalized density matrices is embedded in a
real affine space of dimension

\begin{equation}
d^2-1=(d_A d_B)^2-1 .
\end{equation}

We denote the set of all quantum states by

\begin{equation}
\mathcal{Q}
=
\left\{
\rho\mid
\rho=\rho^\dagger,\;
\rho\geq0,\;
\Tr(\rho)=1
\right\}.
\end{equation}

For a bipartite system, the separable states form the subset

\begin{equation}
\mathcal{S}
=
\left\{
\rho_{AB}
\mid
\rho_{AB}
=
\sum_i p_i
\rho_i^A\otimes\rho_i^B
\right\},
\end{equation}

where

\begin{equation}
p_i\geq0,\qquad
\sum_i p_i=1 .
\end{equation}

Clearly,

\begin{equation}
\mathcal{S}\subset\mathcal{Q}.
\end{equation}

A set $C\subseteq\mathbb{R}^n$ is convex if, for any
$x_1,x_2\in C$ and $\lambda\in[0,1]$,

\begin{equation}
\lambda x_1+(1-\lambda)x_2\in C .
\end{equation}

The quantum-state set $\mathcal{Q}$ is convex. Indeed, for any
$\rho_1,\rho_2\in\mathcal{Q}$,

\begin{equation}
\rho_\lambda
=
\lambda\rho_1+(1-\lambda)\rho_2
\end{equation}

remains Hermitian and positive semidefinite, and

\begin{equation}
\Tr(\rho_\lambda)
=
\lambda\Tr(\rho_1)
+
(1-\lambda)\Tr(\rho_2)
=1 .
\end{equation}

Therefore,

\begin{equation}
\rho_\lambda\in\mathcal{Q}.
\end{equation}

Similarly, the separable-state set $\mathcal{S}$ is convex. If
$\rho_1$ and $\rho_2$ are separable, then

\begin{align}
\lambda\rho_1+(1-\lambda)\rho_2
={}&
\lambda
\sum_i p_i
\rho_i^A\otimes\rho_i^B
\nonumber\\
&+
(1-\lambda)
\sum_j q_j
\sigma_j^A\otimes\sigma_j^B ,
\end{align}

which is again a convex decomposition into product states.
Consequently,

\begin{equation}
\lambda\rho_1+(1-\lambda)\rho_2\in\mathcal{S}.
\end{equation}

The maximally mixed state is

\begin{equation}
\rho_0=\frac{I_d}{d}.
\end{equation}

It is strictly positive and therefore lies in the interior of
$\mathcal{Q}$. It is also separable, since

\begin{equation}
\frac{I_d}{d}
=
\frac{I_{d_A}}{d_A}
\otimes
\frac{I_{d_B}}{d_B}.
\end{equation}

Moreover, the maximally mixed state is an interior point of the
separable-state set. Hence,

\begin{equation}
\rho_0
\in
\operatorname{int}(\mathcal{S})
\subset
\operatorname{int}(\mathcal{Q}).
\end{equation}

We therefore choose $\rho_0$ as the origin of the state-space
representation.

The two convex sets are not separated by their boundaries. In
particular, there exist states which belong to both boundaries,

\begin{equation}
\partial\mathcal{S}
\cap
\partial\mathcal{Q}
\neq
\varnothing .
\end{equation}

Rank-deficient separable states provide examples of such points.
Thus, the geometry relevant to the following discussion can be
summarized as

\begin{equation}
\boxed{
\mathcal{S}
\subset
\mathcal{Q}
\subset
\mathbb{R}^{(d_A d_B)^2-1}.
}
\end{equation}

In the two-qubit case, $d_A=d_B=2$, and the state space is therefore
a convex subset of $\mathbb{R}^{15}$.

%========================================================
\section{Geometric Quantities}
%========================================================

Let $\rho$ be an entangled state and let

\begin{equation}
\rho_0=\frac{I_d}{d}
\end{equation}

denote the maximally mixed state\cite{GurvitsBarnum2002,GurvitsBarnum2003}. We use $\rho_0$ as the reference
state and consider the ray connecting $\rho_0$ and $\rho$.

Let $\rho_{\mathrm{sep}}^{*}$ denote the state at which this ray
intersects the boundary of the separable-state set, and let
$\rho_{\mathrm{q}}^{*}$ denote the state at which the same ray
intersects the boundary of the quantum-state set. Thus,

\begin{equation}
\rho_0
\longrightarrow
\rho_{\mathrm{sep}}^{*}
\longrightarrow
\rho
\longrightarrow
\rho_{\mathrm{q}}^{*}.
\end{equation}

Using the vector representation introduced above, the corresponding
geometric distances are defined by

\begin{equation}
\left\|
\mathbf{x}(\rho_{\mathrm{sep}}^{*})
-
\mathbf{x}(\rho_0)
\right\|,
\end{equation}

and

\begin{equation}
\left\|
\mathbf{x}(\rho_{\mathrm{q}}^{*})
-
\mathbf{x}(\rho_0)
\right\|,
\end{equation}

where $\|\cdot\|$ denotes the Euclidean norm.

We therefore define

\begin{equation}
L(\rho)
=
\frac{
\left\|
\mathbf{x}(\rho_{\mathrm{sep}}^{*})
-
\mathbf{x}(\rho_0)
\right\|
}{
\left\|
\mathbf{x}(\rho_{\mathrm{q}}^{*})
-
\mathbf{x}(\rho_0)
\right\|
},
\qquad
0<L(\rho)<1 .
\label{eq:Lrho}
\end{equation}

Since $\rho_0$, $\rho_{\mathrm{sep}}^{*}$, $\rho$, and
$\rho_{\mathrm{q}}^{*}$ lie on the same ray, the distance between the
two boundary points is

\begin{equation}
\left\|
\mathbf{x}(\rho_{\mathrm{q}}^{*})
-
\mathbf{x}(\rho_{\mathrm{sep}}^{*})
\right\|
=
\left\|
\mathbf{x}(\rho_{\mathrm{q}}^{*})
-
\mathbf{x}(\rho_0)
\right\|
-
\left\|
\mathbf{x}(\rho_{\mathrm{sep}}^{*})
-
\mathbf{x}(\rho_0)
\right\|.
\end{equation}

We therefore define the geometric entangled-space quantity

\begin{equation}
G(\rho)
=
\frac{
\left\|
\mathbf{x}(\rho_{\mathrm{q}}^{*})
-
\mathbf{x}(\rho_{\mathrm{sep}}^{*})
\right\|
}{
\left\|
\mathbf{x}(\rho_{\mathrm{sep}}^{*})
-
\mathbf{x}(\rho_0)
\right\|
}
=
\frac{1-L(\rho)}{L(\rho)}.
\label{eq:Grho}
\end{equation}

The quantity $G(\rho)$ describes the size of the entangled region
available along the direction determined by $\rho$.

We next consider mixing $\rho$ with the maximally mixed state\cite{VidalTarrach1999},

\begin{equation}
\rho(p)
=
p\rho+(1-p)\rho_0,
\end{equation}

where $p$ is the largest value for which $\rho(p)$ remains
separable. At the critical value of $p$, the state reaches the
separable-state boundary, so that

\begin{equation}
p
=
\frac{
\left\|
\mathbf{x}(\rho_{\mathrm{sep}}^{*})
-
\mathbf{x}(\rho_0)
\right\|
}{
\left\|
\mathbf{x}(\rho)
-
\mathbf{x}(\rho_0)
\right\|
}.
\end{equation}

Therefore,

\begin{equation}
\left\|
\mathbf{x}(\rho)
-
\mathbf{x}(\rho_{\mathrm{sep}}^{*})
\right\|
=
\left\|
\mathbf{x}(\rho)
-
\mathbf{x}(\rho_0)
\right\|
-
\left\|
\mathbf{x}(\rho_{\mathrm{sep}}^{*})
-
\mathbf{x}(\rho_0)
\right\|,
\end{equation}

and hence

\begin{equation}
R_{\mathrm{mix}}(\rho)
=
\frac{
\left\|
\mathbf{x}(\rho)
-
\mathbf{x}(\rho_{\mathrm{sep}}^{*})
\right\|
}{
\left\|
\mathbf{x}(\rho_{\mathrm{sep}}^{*})
-
\mathbf{x}(\rho_0)
\right\|
}
=
\frac{1-p}{p}.
\label{eq:Rmixrho}
\end{equation}

We then define the relative entanglement degree as the ratio between
the robustness-related quantity and the geometric entangled-space
quantity,

\begin{equation}
Q(\rho)
=
\frac{R_{\mathrm{mix}}(\rho)}
{G(\rho)}.
\label{eq:Qrho}
\end{equation}

Thus,

\begin{equation}
\boxed{
Q(\rho)
=
\frac{(1-p)/p}
{(1-L(\rho))/L(\rho)}
=
\frac{(1-p)L(\rho)}
{p[1-L(\rho)]}.
}
\end{equation}

Here, $G(\rho)$ describes the total entangled space along the
direction of $\rho$, while $R_{\mathrm{mix}}(\rho)$ describes the
actual entangled distance of $\rho$ from the separable-state boundary.
The quantity $Q(\rho)$ compares these two quantities.

%========================================================
\section{Best Separable Approximation}
%========================================================

For a mixed entangled state $\rho$, the Best Separable Approximation
(BSA) decomposes the state into a separable part and a remaining
entangled part\cite{Lewenstein1998},

\begin{equation}
\rho
=
(1-p_0)\rho_A
+
p_0\rho_B ,
\label{eq:BSA}
\end{equation}

where $\rho_A$ is a separable state, $\rho_B$ is an entangled state,
and $p_0$ is chosen such that the separable weight $1-p_0$ is maximal\cite{Lewenstein1998,KarnasLewenstein2001,Thiang2010}.

In the geometric representation, we denote the target state $\rho$ by
the point $R$, the separable component $\rho_A$ by the point $A$, and
the entangled component $\rho_B$ by the point $B$. Therefore, the BSA
decomposition can be written geometrically as

\begin{equation}
\overrightarrow{OR}
=
(1-p_0)\overrightarrow{OA}
+
p_0\overrightarrow{OB}.
\label{eq:vector_BSA}
\end{equation}

It follows immediately that the four points $O$, $A$, $B$, and $R$
lie in a common two-dimensional plane. Indeed, the vector
$\overrightarrow{OR}$ is a linear combination of
$\overrightarrow{OA}$ and $\overrightarrow{OB}$. When
$\overrightarrow{OA}$ and $\overrightarrow{OB}$ are not collinear,
they span a unique plane containing $O$, $A$, $B$, and $R$.

When $\overrightarrow{OA}$ and $\overrightarrow{OB}$ are collinear,
the four points already lie on a single line. In this case, we may
choose any plane containing this line, and the following geometric
construction remains unchanged.

Therefore, although the original state space has dimension
$(d_A d_B)^2-1$, the geometric relations associated with the BSA
decomposition can always be represented in a two-dimensional section.
An illustration of this planar section is shown in Fig.~1.

\begin{figure}[htbp]
\centering
\includegraphics[width=0.70\linewidth]{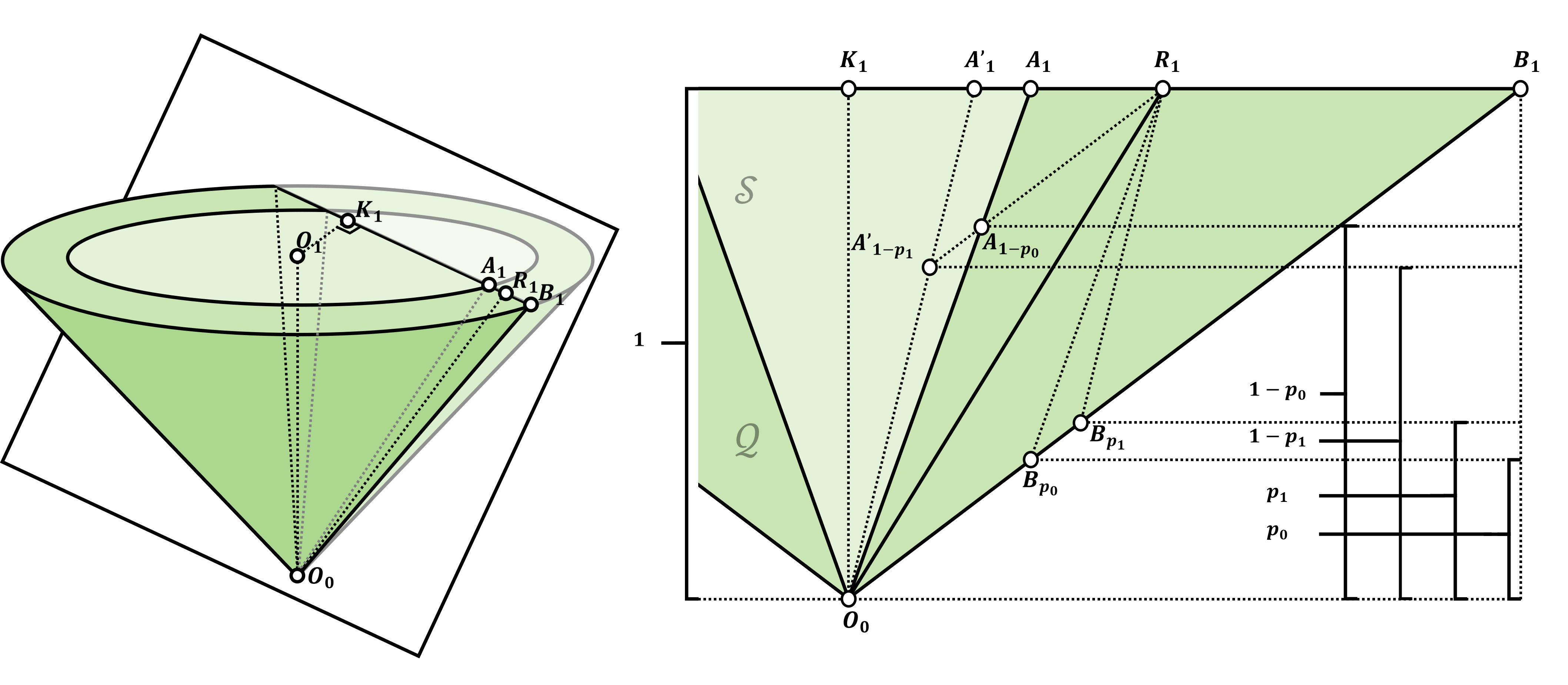}
\caption{
A two-dimensional section of the high-dimensional quantum-state
geometry containing the points $O$, $A$, $B$, and $R$. Since
$\overrightarrow{OR}$ is a convex combination of
$\overrightarrow{OA}$ and $\overrightarrow{OB}$, these points lie in a
common plane. The higher-dimensional convex geometry can therefore be
studied in this plane without changing the geometric relations used
below.
}
\end{figure}

%--------------------------------------------------------
\subsection{The Entangled Component Lies on the Quantum-State Boundary}
%--------------------------------------------------------

We first show that the entangled component of an optimal BSA must lie
on the boundary of the quantum-state set.

Suppose, on the contrary, that $\rho_B$ is an interior point of
$\mathcal{Q}$. Since $\rho_B$ is an interior point, there exists a
sufficiently small $\epsilon>0$ such that

\begin{equation}
\rho_B-\epsilon\rho_0\geq0,
\end{equation}

where

\begin{equation}
\rho_0=\frac{I_d}{d}
\end{equation}

is the maximally mixed state.

Because
$\Tr(\rho_B)=\Tr(\rho_0)=1$, the normalized operator

\begin{equation}
\tilde{\rho}_B
=
\frac{\rho_B-\epsilon\rho_0}{1-\epsilon}
\end{equation}

satisfies

\begin{equation}
\tilde{\rho}_B
=
\tilde{\rho}_B^\dagger,
\qquad
\tilde{\rho}_B\geq0,
\qquad
\Tr(\tilde{\rho}_B)=1.
\end{equation}

Thus, $\tilde{\rho}_B$ is still a quantum state, and

\begin{equation}
\rho_B
=
\epsilon\rho_0
+
(1-\epsilon)\tilde{\rho}_B .
\end{equation}

The maximally mixed state is separable\cite{GurvitsBarnum2002} because

\begin{equation}
\rho_0
=
\frac{I_{d_A}}{d_A}
\otimes
\frac{I_{d_B}}{d_B}.
\end{equation}

Substituting the above decomposition of $\rho_B$ into Eq.~\eqref{eq:BSA}
gives

\begin{align}
\rho
={}&
(1-p_0)\rho_A
+
p_0\epsilon\rho_0
+
p_0(1-\epsilon)\tilde{\rho}_B .
\end{align}

Since both $\rho_A$ and $\rho_0$ are separable, they can be combined
into a new separable state $\tilde{\rho}_A$,

\begin{equation}
\tilde{\rho}_A
=
\frac{
(1-p_0)\rho_A+p_0\epsilon\rho_0
}{
(1-p_0)+p_0\epsilon
}.
\end{equation}

Therefore,

\begin{equation}
\rho
=
\left[(1-p_0)+p_0\epsilon\right]\tilde{\rho}_A
+
p_0(1-\epsilon)\tilde{\rho}_B .
\end{equation}

The new separable weight is

\begin{equation}
(1-p_0)+p_0\epsilon
>
1-p_0 ,
\end{equation}

which contradicts the optimality of the BSA decomposition.

Hence,

\begin{equation}
\boxed{
\rho_B\in\partial\mathcal{Q}.
}
\label{eq:B_boundary}
\end{equation}

%--------------------------------------------------------
\subsection{The Separable Component Lies on the Separable-State Boundary}
%--------------------------------------------------------

We next show that the separable component of the optimal BSA
decomposition must lie on the boundary of the separable-state set.

To make the decomposition transparent geometrically, we introduce an
additional coordinate $Z$. For a state represented by the vector
$\mathbf{x}$ in the original state space, its point in the extended
space is written as

\begin{equation}
(Z,Z\mathbf{x}).
\end{equation}

Thus, the original state space is located at $Z=1$, while decreasing
$Z$ corresponds to a uniform contraction of the state-space geometry
toward the origin.

For the BSA decomposition

\begin{equation}
\rho
=
(1-p_0)\rho_A+p_0\rho_B,
\end{equation}

the separable and entangled components are represented by points in
different $Z$-layers. In particular, the target state $R$ is located
at

\begin{equation}
R=(1,\mathbf{r}),
\end{equation}

while a separable component with weight $p'$ and state vector
$\mathbf{a}'$ is represented by

\begin{equation}
A'=(p',p'\mathbf{a}').
\end{equation}

Assume, for contradiction, that the separable component of the
optimal BSA decomposition is an interior point of the separable
convex body. Denote this point by $A'$. Since $A'$ is an interior
point, the line connecting $A'$ to the target point $R$ must intersect
the boundary of the contracted separable convex body at another point,
which we denote by

\begin{equation}
A=(p,p\mathbf{a}).
\end{equation}

Because $A$ lies farther from $A'$ in the direction of $R$, while the
target point has fixed coordinate $Z=1$, its $Z$ coordinate satisfies

\begin{equation}
p\geq p'.
\end{equation}

The points $A'$, $A$, and $R$ are collinear. Hence, replacing $A'$ by
$A$ does not change the direction of the vector connecting the
separable component to the target state. In particular,

\begin{equation}
\overrightarrow{AR}
\parallel
\overrightarrow{A'R}.
\end{equation}

Therefore, if the vector from $A'$ to $R$ represents an allowed
entangled component, the vector from $A$ to $R$ has the same direction
in the original state space and remains an allowed entangled
component under the same geometric construction.

The new point $A$ is still a separable state, but its $Z$ coordinate
satisfies

\begin{equation}
p\geq p'.
\end{equation}

Thus, the corresponding decomposition contains a separable
contribution with weight at least as large as that associated with
$A'$. Since $A'$ was assumed to represent the optimal BSA
decomposition and $A$ lies strictly farther toward the target whenever
$A'$ is an interior point, this gives a contradiction to the maximality
of the separable weight.

Hence, the separable component of the optimal BSA cannot lie in the
interior of the separable convex body. Therefore,

\begin{equation}
\boxed{
\rho_A\in\partial\mathcal{S}.
}
\end{equation}

%========================================================
\section{Geometric Relations for the BSA}
%========================================================

\subsection{Geometric Construction}

We now consider the geometric structure associated with the BSA
decomposition. Let the BSA of an entangled state $\rho$ be written as

\begin{equation}
\rho
=
(1-p_0)\rho_A+p_0\rho_B ,
\end{equation}

where $\rho_A$ is the separable component and $\rho_B$ is the
entangled component.

We denote the target state $\rho$ by the point $R$, the separable
component $\rho_A$ by the point $A$, and the entangled component
$\rho_B$ by the point $B$. From the BSA decomposition,

\begin{equation}
\overrightarrow{OR}
=
(1-p_0)\overrightarrow{OA}
+
p_0\overrightarrow{OB}.
\end{equation}

Therefore, the four points $O$, $A$, $B$, and $R$ lie in the same
two-dimensional subspace. In the non-collinear case,
$\overrightarrow{OA}$ and $\overrightarrow{OB}$ determine this
subspace. If they are collinear, any plane containing the line $OB$
can be chosen.

All the following geometric constructions are performed within this
two-dimensional subspace, as illustrated in Fig.~2.

\begin{figure}[htbp]
\centering
\includegraphics[width=0.78\linewidth]{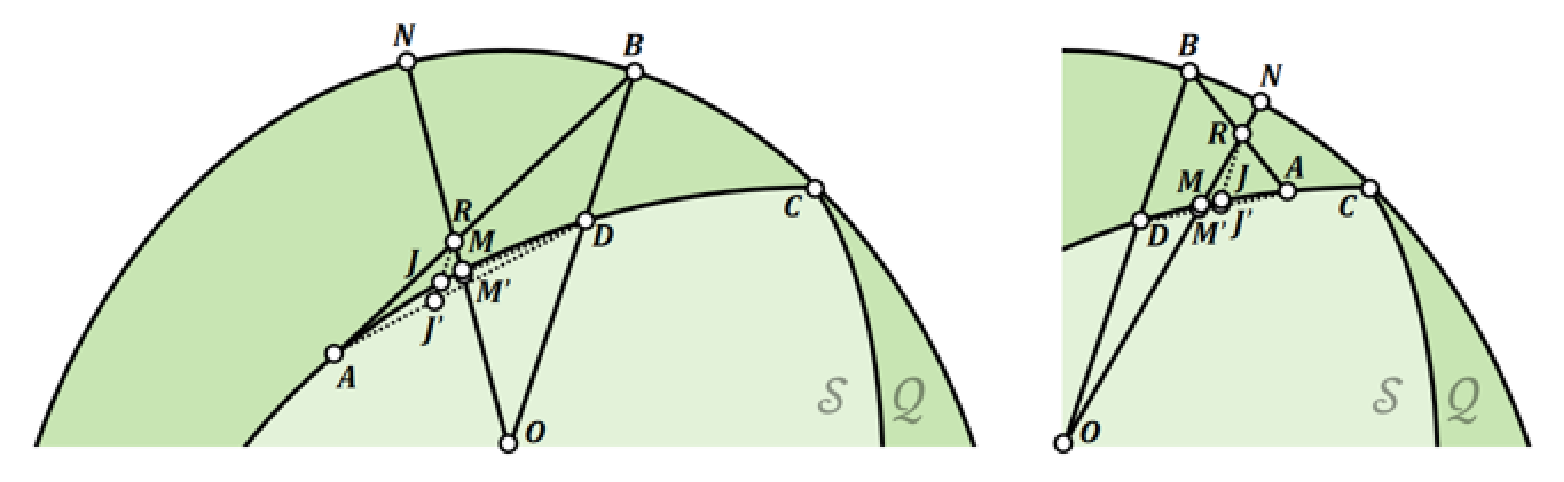}
\caption{
Geometric construction for the BSA decomposition. The points $O$,
$A$, $B$, and $R$ lie in a common two-dimensional subspace. The point
$A$ represents the separable component of the BSA, $B$ the entangled
component, and $R$ the target state. The points $D$, $M$, and $N$ are
the intersections of the rays $OB$ and $OR$ with the boundaries of
the separable-state and quantum-state sets. The point $C$ belongs to
the intersection of the two boundaries within the chosen
two-dimensional subspace. The auxiliary points $J$, $J'$, and $M'$
are introduced for the subsequent geometric derivation.
}
\end{figure}

According to the boundary properties established in the previous
section,

\begin{equation}
A\in\partial\mathcal{S},
\qquad
B\in\partial\mathcal{Q}.
\end{equation}

The point $D$ is defined as the intersection of the ray $OB$ with the
boundary of the separable-state set,

\begin{equation}
D\in\partial\mathcal{S},
\qquad
O,D,B\ \text{are collinear}.
\end{equation}

The ray $OR$ intersects the boundary of the separable-state set at
the point $M$,

\begin{equation}
M\in\partial\mathcal{S},
\qquad
O,M,R\ \text{are collinear},
\end{equation}

and intersects the boundary of the quantum-state set at the point
$N$,

\begin{equation}
N\in\partial\mathcal{Q},
\qquad
O,R,N\ \text{are collinear}.
\end{equation}

Thus, along the direction $OR$,

\begin{equation}
O-M-R-N.
\end{equation}

Within the chosen two-dimensional subspace, the boundaries of the
separable-state set and the quantum-state set may have common points.
We denote such a boundary-intersection point by $C$:

\begin{equation}
C\in
\partial\mathcal{S}\cap\partial\mathcal{Q}.
\end{equation}

The point $C$ is defined only by the intersection of the two
boundaries in the chosen two-dimensional subspace and does not rely on
any additional supporting hyperplane construction.

For the subsequent geometric relations, draw through the target point
$R$ a line parallel to $BO$. Let its intersection with the separable
boundary be denoted by $J$,

\begin{equation}
RJ\parallel BO,
\qquad
J\in\partial\mathcal{S}.
\end{equation}Since
\begin{equation}
\overrightarrow{OR}
=
(1-p_0)\overrightarrow{OA}
+
p_0\overrightarrow{OB},
\end{equation}
we have
\begin{equation}
\overrightarrow{OR}
-
p_0\overrightarrow{OB}
=
(1-p_0)\overrightarrow{OA}.
\end{equation}
Because $A$ lies on the boundary of the separable-state convex set,
the point corresponding to $(1-p_0)\overrightarrow{OA}$ lies inside
the separable-state convex set. Thus, starting from the entangled point
$R$ and moving along the direction parallel to $BO$, the ray starts
outside the separable-state convex set and subsequently enters it.
Therefore, it must intersect the separable-state boundary at a point
$J$.

The same line intersects the line segment $AD$ at the point $J'$,

\begin{equation}
J'\in AD,
\qquad
R,J,J'\ \text{are collinear}.
\end{equation}

Finally, connect $J$ and $D$. Their connecting line intersects the
line $OM$ at the point $M'$,

\begin{equation}
M'\in OM,
\qquad
M'\in JD.
\end{equation}

The geometric construction is therefore characterized by the points
\[
O,\ A,\ B,\ R,\ C,\ D,\ M,\ N,\ J,\ J',\ M',
\]
together with the collinearities and boundary relations defined above.
The next section uses these points to derive the relations between the
BSA weight, the geometric quantity, and the robustness.

%--------------------------------------------------------
\subsection{Geometric Relations and BSA Bounds}
%--------------------------------------------------------

We now derive the relations between the BSA coefficient, the
robustness with respect to the maximally mixed state\cite{Lewenstein1998,VidalTarrach1999}, and the
geometric quantities introduced above.

Since the target state $R$ is the convex combination of $A$ and $B$,
we have

\begin{equation}
\overrightarrow{OR}
=
(1-p_0)\overrightarrow{OA}
+
p_0\overrightarrow{OB}.
\end{equation}

Because $A$, $R$, and $B$ are collinear, this gives

\begin{equation}
\frac{AR}{AB}=p_0.
\label{eq:AR_AB}
\end{equation}

We now consider the line $OR$, which intersects the separable-state
and quantum-state boundaries at $M$ and $N$, respectively. Since $M$, $R$, and $N$ are collinear, the target state can also be
written as a convex combination of the states represented by $M$ and
$N$,

\begin{equation}
\overrightarrow{OR}
=
(1-q)\overrightarrow{OM}
+
q\overrightarrow{ON},
\end{equation}

where

\begin{equation}
q
=
\frac{MR}{MN}.
\end{equation}

where

\begin{equation}
q
=
\frac{MR}{MN}.
\end{equation}

Thus, the segment $MRN$ provides another feasible decomposition of the
target state, with entangled weight $q$. In contrast, $p_0$ is the
entangled weight of the optimal BSA decomposition. By the definition
of the BSA, its entangled weight cannot be larger than that of any
other feasible decomposition. Hence,

\begin{equation}
p_0
\leq
q
=
\frac{MR}{MN}.
\label{eq:p0_MRN}
\end{equation}

From the definitions of $G(\rho)$ and
$R_{\mathrm{mix}}(\rho)$,

\begin{equation}
G(\rho)
=
\frac{MN}{OM},
\end{equation}

and

\begin{equation}
R_{\mathrm{mix}}(\rho)
=
\frac{MR}{OM}.
\end{equation}

Therefore,

\begin{equation}
Q(\rho)
=
\frac{R_{\mathrm{mix}}(\rho)}
{G(\rho)}
=
\frac{MR}{MN}
=
\frac{1-p}{p}
\frac{L_R}{1-L_R}.
\end{equation}

Combining this relation with Eq.~\eqref{eq:p0_MRN}, we obtain

\begin{equation}
\boxed{
Q(\rho)\geq p_0.
}
\label{eq:Q_p0}
\end{equation}

Since the BSA separable weight is $1-p_0$, Eq.~\eqref{eq:Q_p0}
immediately gives the lower bound

\begin{equation}
\boxed{
1-Q(\rho)\leq 1-p_0.
}
\label{eq:BSA_lower}
\end{equation}

Thus, $1-Q(\rho)$ provides a geometric lower bound for the separable
weight in the BSA decomposition.

We next derive the relation between the geometric entangled-space
sizes of the target state $R$ and the entangled component $B$.

Consider the auxiliary points $J$, $J'$ and $M'$ introduced in the
previous subsection. Since

\begin{equation}
RJ\parallel BO
\end{equation}

and

\begin{equation}
J'\in AD,
\end{equation}

the corresponding triangles are similar, which gives

\begin{equation}
\frac{RJ'}{BD}
=
\frac{AR}{AB}
=
p_0.
\label{eq:RJ_BD}
\end{equation}

Moreover, since $M'\in JD$ and $M'\in OR$, the corresponding triangles
give

\begin{equation}
\frac{M'R}{OM'}
=
\frac{RJ}{OD}.
\label{eq:similarity}
\end{equation}

Because $M'$ lies inside the separable convex body on the direction
$OM$, while $M$ is its boundary point, we have

\begin{equation}
OM'\leq OM.
\end{equation}

At the same time, since $M'$ lies between the origin and $R$,

\begin{equation}
M'R\geq MR.
\end{equation}

Therefore,

\begin{equation}
\frac{M'R}{OM'}
\geq
\frac{MR}{OM}.
\end{equation}

Using

\begin{equation}
\frac{MR}{OM}
=
\frac{1-p}{p},
\end{equation}

we obtain from Eq.~\eqref{eq:similarity}

\begin{equation}
RJ
\geq
\frac{1-p}{p}\,OD.
\label{eq:RJ_lower}
\end{equation}

Since $J'$ lies on the line segment $AD$, while $J$ is on the
corresponding line through $R$, the construction gives

\begin{equation}
RJ'\geq RJ.
\end{equation}

Together with Eq.~\eqref{eq:RJ_BD}, this yields

\begin{equation}
p_0BD
\geq
\frac{1-p}{p}\,OD.
\end{equation}

Hence,

\begin{equation}
p_0
\geq
\frac{1-p}{p}
\frac{OD}{BD}.
\label{eq:p0_general}
\end{equation}

The ratio on the right-hand side can be written in terms of the
geometric parameter of the state represented by $B$. Since $D$ and $B$
lie on the same radial direction,

\begin{equation}
L_B=\frac{OD}{OB},
\end{equation}

and therefore

\begin{equation}
\frac{BD}{OD}
=
\frac{1-L_B}{L_B}.
\end{equation}

Thus,

\begin{equation}
\frac{OD}{BD}
=
\frac{L_B}{1-L_B}.
\end{equation}

Substituting this into Eq.~\eqref{eq:p0_general}, we obtain

\begin{equation}
p_0
\geq
\frac{1-p}{p}
\frac{L_B}{1-L_B}.
\label{eq:p0_LB}
\end{equation}

Combining Eqs.~\eqref{eq:Q_p0} and~\eqref{eq:p0_LB}, we obtain the
two-sided bound for the entangled weight $p_0$ in the optimal BSA
decomposition,

\begin{equation}
\boxed{
\frac{1-p}{p}
\frac{L_B}{1-L_B}
\leq
p_0
\leq
Q(\rho)=\frac{1-p}{p}
\frac{L_R}{1-L_R}.
}
\label{eq:p0_bounds}
\end{equation}

Equivalently, since the separable weight of the BSA is

\begin{equation}
p_{\mathrm{BSA}}=1-p_0,
\end{equation}

we obtain the corresponding two-sided bound

\begin{equation}
\boxed{
1-Q(\rho)
\leq
p_{\mathrm{BSA}}
\leq
1-
\frac{1-p}{p}
\frac{L_B}{1-L_B}.
}
\label{eq:BSA_bounds}
\end{equation}

Therefore, the geometric quantity $Q(\rho)$ provides an upper bound
for the entangled weight and, equivalently, a lower bound for the
separable weight in the optimal BSA decomposition. The quantity
$L_B$ gives the complementary bound from the geometric size of the
entangled component.

In addition, the geometric construction gives the ordering relation

\begin{equation}
\boxed{
\frac{1-L_R}{L_R}
\leq
\frac{1-L_B}{L_B}.
}
\label{eq:geometric_order}
\end{equation}This relation shows that the entangled component extracted by the
optimal BSA decomposition lies in a direction with an entangled-space
size no smaller than that of the original mixed state.

Thus, the entangled component obtained from the optimal BSA
decomposition has an entangled-space size no smaller than that of the
original mixed state.
These relations depend only on the convex geometry of the two nested
state sets and the optimality of the BSA decomposition.

%========================================================
\section{Two-Qubit Case}
%========================================================

For two-qubit states, the PPT criterion provides a necessary and
sufficient condition for separability\cite{Peres1996,Horodecki1996PPT}. This allows the geometric
quantities introduced above to be obtained directly from the density
matrix.

Let $\rho$ be the target state and define

\begin{equation}
r
=
\rho-\frac{I_4}{4}.
\end{equation}

A general point on the ray starting from the maximally mixed state and
passing through $\rho$ can then be written as

\begin{equation}
K(k)
=
\frac{I_4}{4}+kr .
\label{eq:K}
\end{equation}

At $k=1$,

\begin{equation}
K(1)=\rho.
\end{equation}

The first point at which this ray reaches the boundary of the
quantum-state set is determined by

\begin{equation}
\boxed{
k_B
=
\min
\left\{
k\geq1
\;\middle|\;
\det K(k)=0
\right\}.
}
\label{eq:kB}
\end{equation}

Similarly, taking the partial transpose of $r$,

\begin{equation}
r'
=
r^{T_B},
\end{equation}

we define

\begin{equation}
K'(k)
=
\frac{I_4}{4}+kr'
=
\frac{I_4}{4}
+
k\left(
\rho-\frac{I_4}{4}
\right)^{T_B}.
\label{eq:Kprime}
\end{equation}

Since PPT is equivalent to separability for two-qubit states\cite{Horodecki1996PPT}, the
first point at which the ray reaches the separable-state boundary is

\begin{equation}
\boxed{
k_A
=
\min
\left\{
0<k\leq1
\;\middle|\;
\det K'(k)=0
\right\}.
}
\label{eq:kA}
\end{equation}

Because the radial distance is proportional to $k$, the geometric
parameter of the target state is

\begin{equation}
L_R
=
\frac{k_A}{k_B}.
\label{eq:LR_k}
\end{equation}

Therefore,

\begin{equation}
\boxed{
G(R)
=
\frac{1-L_R}{L_R}
=
\frac{k_B-k_A}{k_A}.
}
\label{eq:GR_k}
\end{equation}

On the other hand, mixing the target state with the maximally mixed
state gives

\begin{equation}
\rho(p)
=
\frac{I_4}{4}
+
p\left(
\rho-\frac{I_4}{4}
\right).
\end{equation}

The largest value of $p$ for which the state remains separable is
precisely $k_A$. Thus,

\begin{equation}
p=k_A,
\end{equation}

and the corresponding robustness quantity is\cite{VidalTarrach1999}

\begin{equation}
\boxed{
R_{\mathrm{mix}}(\rho)
=
\frac{1-p}{p}
=
\frac{1-k_A}{k_A}.
}
\label{eq:Rmix_k}
\end{equation}

Consequently, the relative entanglement degree can be written as

\begin{equation}
\boxed{
Q(\rho)
=
\frac{R_{\mathrm{mix}}(\rho)}{G(\rho)}
=
\frac{1-k_A}{k_B-k_A}.
}
\label{eq:Q_k}
\end{equation}

%--------------------------------------------------------
\subsection{Pure Entangled Component in the Case}
%--------------------------------------------------------

We next consider the entangled component $\rho_B$ in the optimal BSA
decomposition.

For a two-qubit state with rank larger than one, its support has
dimension at least two. Any two-dimensional subspace of a two-qubit
Hilbert space contains a product vector
\cite{Horodecki2000LowRank}. Therefore, if an entangled component
$\rho_B$ satisfies

\begin{equation}
\operatorname{rank}(\rho_B)\geq2 ,
\end{equation}

its support contains a separable pure state. Consequently, a nonzero
separable contribution can be extracted from $\rho_B$ while keeping the
remaining operator Hermitian and positive semidefinite.

Hence, an entangled component in an optimal BSA decomposition cannot
have rank larger than one, since otherwise a further separable
contribution could be extracted and the separable weight could be
increased, contradicting the optimality of the BSA.

Therefore, for the two-qubit case, the remaining entangled
component in the optimal decomposition must be

\begin{equation}
\boxed{
\rho_B
=
|\psi_B\rangle\langle\psi_B|
}
\end{equation}

a pure entangled state.

Using local unitary transformations, an arbitrary two-qubit pure
entangled state can be written in Schmidt form as

\begin{equation}
|\psi_B\rangle
=
\cos\theta\,|00\rangle
+
\sin\theta\,|11\rangle ,
\end{equation}

where

\begin{equation}
0<\theta\leq\frac{\pi}{4}.
\end{equation}

%--------------------------------------------------------
\subsection{Geometric Size of the Pure Entangled Component}
%--------------------------------------------------------

Consider the ray from the maximally mixed state toward the pure
entangled state $\rho_B$. The corresponding states are

\begin{equation}
\rho_B(k)
=
\frac{I_4}{4}
+
k
\left(
|\psi_B\rangle\langle\psi_B|
-
\frac{I_4}{4}
\right).
\end{equation}

Using the PPT criterion, the separable-state boundary is reached at\cite{Peres1996,Horodecki1996PPT}

\begin{equation}
k_A^{(B)}
=
\frac{1}
{1+2|\sin 2\theta|}.
\end{equation}

Since $\rho_B$ is pure, it lies on the boundary of the quantum-state
set, and therefore

\begin{equation}
k_B^{(B)}=1.
\end{equation}

Hence, the corresponding geometric parameter is

\begin{equation}
L_B
=
\frac{k_A^{(B)}}{k_B^{(B)}}
=
\frac{1}
{1+2|\sin2\theta|}.
\end{equation}

It follows that

\begin{equation}
\boxed{
\frac{1-L_B}{L_B}
=
2|\sin2\theta|.
}
\label{eq:GB_pure}
\end{equation}

Equivalently,

\begin{equation}
\boxed{
\frac{L_B}{1-L_B}
=
\frac{1}{2|\sin2\theta|}.
}
\label{eq:LB_ratio}
\end{equation}

Since

\begin{equation}
|\sin2\theta|\leq1,
\end{equation}

the above quantity satisfies

\begin{equation}
\boxed{
\frac{L_B}{1-L_B}\geq\frac12.
}
\end{equation}

The minimum value $1/2$ is attained when

\begin{equation}
\theta=\frac{\pi}{4},
\end{equation}

corresponding to a maximally entangled state. Thus, among two-qubit
pure entangled states, the maximally entangled state has the largest
geometric entangled-space size,

\begin{equation}
\frac{1-L_B}{L_B}\leq2.
\end{equation}

%--------------------------------------------------------
\subsection{Two-Qubit BSA Bounds}
%--------------------------------------------------------

The general geometric relation derived above gives

\begin{equation}
\frac{1-p}{p}
\frac{L_B}{1-L_B}
\leq
p_0
\leq
Q(\rho).
\end{equation}

Using

\begin{equation}
\frac{L_B}{1-L_B}
\geq
\frac12,
\end{equation}

we obtain the two-qubit bound

\begin{equation}
\boxed{
\frac{1-p}{2p}
\leq
p_0
\leq
Q(\rho).
}
\label{eq:twoqubit_p0}
\end{equation}

Since the BSA separable weight is

\begin{equation}
p_{\mathrm{BSA}}
=
1-p_0,
\end{equation}

the corresponding two-sided bound is

\begin{equation}
\boxed{
1-Q(\rho)
\leq
p_{\mathrm{BSA}}
\leq
1-\frac{1-p}{2p}.
}
\label{eq:twoqubit_BSA}
\end{equation}

Thus, for a entangled two-qubit state, the geometric quantity
$Q(\rho)$ gives an upper bound on the entangled weight of the optimal
BSA, while the PPT analysis of the pure entangled component provides
the corresponding lower bound.

%--------------------------------------------------------
\subsection{Summary of the Two-Qubit Results}
%--------------------------------------------------------

The results obtained above give the following two-sided bound for the
BSA entangled weight of a two-qubit state,

\begin{equation}
\frac{1-p}{2p}
\leq
p_0
\leq
Q(\rho).
\end{equation}

The corresponding bound for the separable weight is

\begin{equation}
1-Q(\rho)
\leq
p_{\mathrm{BSA}}
\leq
1-\frac{1-p}{2p}.
\end{equation}

The upper bound $p_0\leq Q(\rho)$ becomes an equality when all points
on the line segment $AD$ belong to the boundary of the separable-state
convex set. In this case, the geometric construction gives directly

\begin{equation}
p_0=Q(\rho).
\end{equation}

For the lower bound, the equality
\begin{equation}
p_0=\frac{1-p}{2p}
\end{equation}
is attained when the entangled component is maximally entangled,
corresponding to $\theta=\pi/4$. In particular, this occurs for the
corresponding Werner state constructed from a maximally entangled
two-qubit state.

Thus, the two-qubit analysis provides explicit cases in which both
sides of the general BSA bound can be saturated.

\section{Conclusion}

In this work, we studied the relation between quantum entanglement and
the convex geometry of bipartite quantum state spaces. The set of
separable states was considered as a convex subset of the full
quantum-state set, with the maximally mixed state chosen as the
reference point.

Based on the radial geometry of these two convex sets, we introduced
the geometric quantity

\begin{equation}
G(\rho)=\frac{1-L(\rho)}{L(\rho)}
\end{equation}

and compared it with the robustness relative to the maximally mixed
state
\cite{VidalTarrach1999},

\begin{equation}
R_{\mathrm{mix}}(\rho)=\frac{1-p}{p}.
\end{equation}

Their ratio defines the relative entanglement degree,

\begin{equation}
Q(\rho)
=
\frac{
R_{\mathrm{mix}}(\rho)
}{
G(\rho)
}.
\end{equation}

We then studied the Best Separable Approximation (BSA)
\cite{Lewenstein1998,KarnasLewenstein2001} and showed that, for an
optimal decomposition, the separable component lies on the boundary
of the separable-state set, while the remaining entangled component
lies on the boundary of the quantum-state set. These properties do not
rely on the PPT criterion or on the special structure of two-qubit
systems.

Using the resulting geometric construction, we obtained the general
relations

\begin{equation}
\frac{1-p}{p}
\frac{L_B}{1-L_B}
\leq
p_0
\leq
Q(\rho),
\end{equation}

and

\begin{equation}
\frac{1-L_R}{L_R}
\leq
\frac{1-L_B}{L_B}.
\end{equation}

The second relation shows that the entangled component obtained from an
optimal BSA has an entangled-space size no smaller than that of the
original mixed state. Thus, the BSA decomposition has a direct
geometric interpretation in terms of the available entangled space.

For two-qubit states, the remaining BSA entangled component is
a pure entangled state, consistent with the special structure
of two-qubit systems
\cite{KarnasLewenstein2001,Horodecki2000LowRank}.
Using the PPT criterion, which is necessary and sufficient for
separability in the two-qubit case
\cite{Peres1996,Horodecki1996PPT},
its geometric parameter can be obtained explicitly, leading to

\begin{equation}
\frac{1-p}{2p}
\leq
p_0
\leq
Q(\rho).
\end{equation}

The corresponding bound for the BSA separable weight is

\begin{equation}
1-Q(\rho)
\leq
p_{\mathrm{BSA}}
\leq
1-\frac{1-p}{2p}.
\end{equation}

The equality conditions discussed above show that both sides of the
two-qubit bound can be saturated for appropriate states.

The main purpose of this work is to provide a simple geometric
description of BSA and its relation to other entanglement quantities.
In higher dimensions, the same geometric relations remain meaningful,
although the explicit determination of the separable boundary is more
difficult because the PPT criterion is no longer sufficient for
separability
\cite{Horodecki1996PPT,Doherty2004,Thiang2010}.

\section*{Acknowledgments}
The author gratefully acknowledges the assistance of OpenAI's GPT models in the writing and preparation of this manuscript, including language polishing and LaTeX typesetting support.

%========================================================
% References
%========================================================

\end{document}